\documentclass[aip,amsmath,amssymb,reprint,graphicx]{revtex4-1}
\usepackage{graphicx}
\usepackage{dcolumn}
\usepackage{bm}
\usepackage[utf8]{inputenc}
\usepackage[T1]{fontenc}
\usepackage{mathptmx}
\usepackage{etoolbox}
\usepackage{upgreek}
\usepackage{hyperref}
\hypersetup{colorlinks=true, linkcolor=black, citecolor=black, urlcolor=black}
\usepackage{tabularx, booktabs}
\usepackage{multirow}

\newcommand{\unit}[1]{\;\mathrm{#1}}

\makeatletter
\def\@email#1#2{%
 \endgroup
 \patchcmd{\titleblock@produce}
  {\frontmatter@RRAPformat}
  {\frontmatter@RRAPformat{\produce@RRAP{*#1\href{mailto:#2}{#2}}}\frontmatter@RRAPformat}
  {}{}
}%
\makeatother

\draft 

\begin{document}

\preprint{AIP/123-QED}

\title{Continuous-Wave Cavity Ring-Down for High-Sensitivity Polarimetry and Magnetometry Measurements} 



\author{Dang-Bao-An Tran}
\affiliation{Department of Chemistry, Physical and Theoretical Chemistry Laboratory, University of Oxford, South Parks Road, Oxford, OX1 3QZ, United Kingdom}
\affiliation{Current Address: Time and Frequency Department, National Physical Laboratory, Teddington, TW11 0LW, United Kingdom}
\author{Evan Edwards}
\author{David P Tew}
\author{Robert Peverall}
\author{Grant A D Ritchie}
    \altaffiliation[Author to whom correspondence should be addressed: ]{\url{grant.ritchie@chem.ox.ac.uk}}
\affiliation{Department of Chemistry, Physical and Theoretical Chemistry Laboratory, University of Oxford, South Parks Road, Oxford, OX1 3QZ, United Kingdom}

\begin{abstract}
We report the development of a novel variant of cavity ring-down polarimetry using a continuous-wave laser operating at $532\unit{nm}$ for highly precise chiroptical activity and magnetometry measurements. The key methodology of the apparatus relies upon the external modulation of the laser frequency at the frequency splitting between non-degenerate left- and right-circularly polarised cavity modes. The method is demonstrated by evaluation of the Verdet constants of crystalline $\text{CeF}_{3}$ and fused silica, in addition to the observation of gas- and solution-phase optical rotations of selected chiral molecules. Specifically, optical rotations of (i) vapours of $\upalpha$-pinene and R-($+$)-limonene, (ii) mutarotating D-glucose in water, and (iii) acidified L-histidine solutions, are determined. The detection sensitivities for the gas- and solution-phase chiral activity measurements are $\sim30\unit{\upmu deg}$ and $\sim120\unit{\upmu deg}$ over a $30\unit{s}$ detection period per cavity roundtrip pass, respectively. Furthermore, the measured optical rotations for R-($+$)-limonene are compared with computations performed using the Turbomole quantum chemistry package. The experimentally observed optically rotatory dispersion of this cyclic monoterpene was thus rationalised via consideration of its room temperature conformer distribution as determined by the aforementioned single-point energy calculations.
\end{abstract}

\maketitle 

\section{Introduction}
\begin{figure*}
    \centering
    \includegraphics[width=0.95\textwidth]{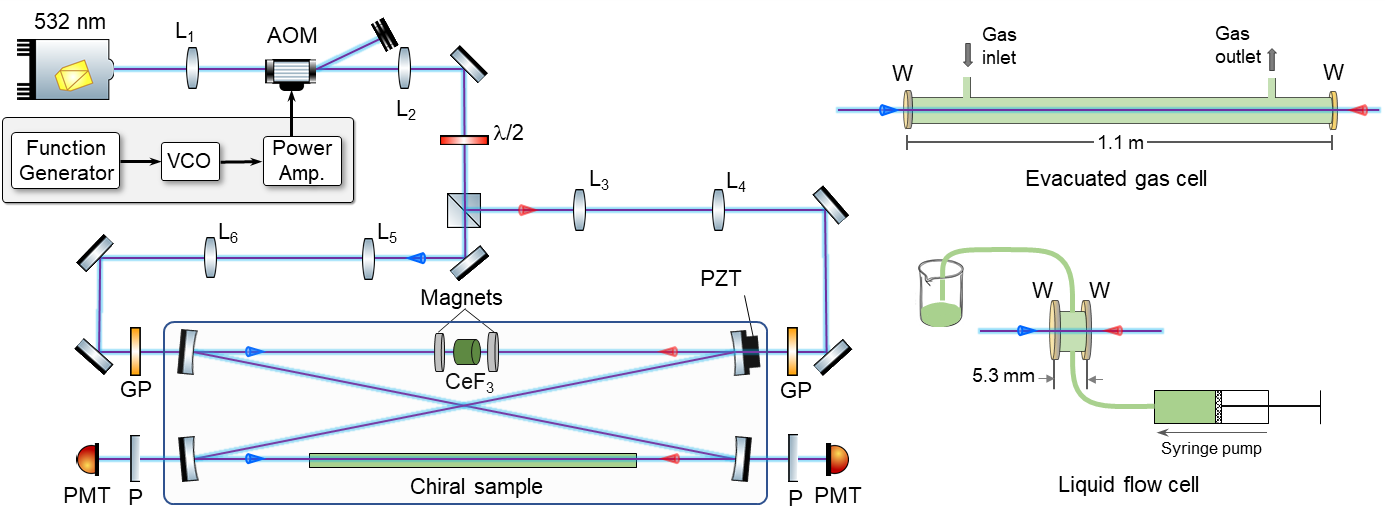}
    \caption{Experimental setup for continuous-wave cavity ring-down polarimetry (CRDP) at $532\unit{nm}$. Light from a diode-pumped solid-state (DPSS) laser at $532\unit{nm}$ is directed to an acousto-optic modulator (AOM). The first-order diffraction beam is separated into clockwise (CW) and counter-clockwise (CCW) beams and then mode matched with the $\text{TEM}_{00}$ mode of a bow-tie cavity comprising four high-reflectivity mirrors (M) ($\text{R}\geq99.99\%)$. A $\text{CeF}_{3}$ crystal is inserted into one arm of the cavity to provide a large bias rotation. For gas-phase optical rotation measurements of chiral samples, a $1.1\unit{m}$ long gas cell is inserted into the other arm of the cavity. For liquid-phase measurements, a $5\unit{mm}$-long flow cell has been used. The CRD traces are detected by two photo-multiplier tubes (PMTs) and the polarisation states of these beams are analysed by two linear polarisers (P). VCO: voltage-controlled oscillator; GP: Glan-Taylor polariser; PZT: piezo-electric actuator; W: window; L; Lens}
    \label{fig:setup}
\end{figure*}
Optical polarimetry is a fundamental technique for studying magnetometry and chirality, and has its basis in the rotation of the polarisation plane of linearly polarised light when passing through a magneto-optic or chiral medium, respectively. Magneto-optic rotational activity plays a key role in technologically important areas such as the development of novel Faraday rotators, magnetometry, and controlling light polarisation\cite{Tamaru2021,Kato2003,Inoue1998,Edwards1995}, while chirality is of crucial importance across many areas of science, from fundamental physics and investigations of parity violation\cite{Cournol2019,Darquie2010}, to pharmacology and biochemistry\cite{Corradini2007}.
\\ \indent Standard laboratory polarimeters have a typical detection precision no better than a few$\unit{mdeg\,Hz^{-1/2}}$ that is limited by birefringence, and are suitable only for liquid-phase observations. It is important to be able to make precise measurements of optical rotation for all phases of matter and thus there has been a continuous effort to develop novel, more sensitive, and precise polarimetric methods\cite{Polavarapu2007,Polavarapu2016,Polavarapu2020}. Particularly attractive in this regard is cavity-based polarimetry, which allows the interaction path length between the light and the magneto-optic/chiral medium to be increased manyfold, thus enhancing the detection sensitivity; for example, cavity ring-down techniques using a high-finesse linear cavity offer a large interaction path length between the light beam and sample of the order of a few km, compared to the few tens of cm afforded by a single cell. However, linear cavities without intracavity optics cannot be employed for chirality observations because the optical rotation is dependent upon the propagation direction of the light beam, and so is suppressed when the linearly polarised light passes back and forth between two cavity mirrors. Vaccaro \textit{et al.} circumvented this problem by developing pulsed laser-based cavity ring-down polarimetry (CRDP) by inserting a pair of intracavity quarter-wave plates to amplify the chiroptical rotation by the number of cavity roundtrip passes---the stress-induced birefringence of the intracavity optics, however, is correspondingly increased\cite{Muller2000,Wilson2005,Craft2021}. This apparatus was used successfully in studying the gas-phase optical rotation of chiral molecules, and utilised a pulsed ns laser source. More recently, Bougas \textit{et al.} demonstrated a continuous-wave (\textit{cw}) variant of CRDP at $408\unit{nm}$ within a linear cavity for which the laser frequency is locked to a cavity resonance. Note, however, that this technique only applies for the measurement of the non-resonant Faraday effect of gases and magneto-optic crystals\cite{Visschers2020}, where the Faraday rotation is independent of the light propagation direction.
\\ \indent Given the limitations associated with using linear cavity geometries, a variety of CRDP instruments based upon bow-tie cavities have recently been demonstrated using pulsed and \textit{cw} lasers. The first pulsed laser CRDP methodology using a bow-tie cavity was demonstrated by Rakitzis and co-workers, and employs two counter-propagating linearly (and orthogonally) polarised beams\cite{Sofikitis2014,Spiliotis2020, Spiliotis2020a,Bougas2015}. An intra-cavity magneto-optic crystal is inserted into one arm of the cavity to provide a large bias rotation angle---the intracavity anisotropies of opposite symmetry are suppressed by a signal reversing technique. The chiroptical rotation is then extracted from the ring-down decay signals of the two counter-propagating beams. These instruments have ring-down times, $\tau$, between 0.5 and $1\unit{\upmu s}$, and have been used for magnetometry measurements and chirality detection across a range of environments, including open air, pressure-controlled vapours, and the liquid phase, with precision $\geq80\unit{\upmu deg}$ per cavity pass. Recently, our group has demonstrated an implementation of bow-tie cavity-enhanced polarimetry with a \textit{cw} laser source at $730\unit{nm}$\cite{Tran2021,Tran2021a}. A key feature of the method is the controlled perturbation of the laser frequency, achieved by the addition of radiofrequency noise to its injection current, which leads to significant improvements in detection sensitivity. In addition, the use of cavity mirrors with a reflectivity $\text{R}\geq99.99\%$, and low-loss intracavity optical components, enables $\tau$ values of $\sim9\unit{\upmu s}$. The detection precision was $\sim10\unit{\upmu deg}$ per cavity round trip for gas-phase optical rotation measurements. A means of achieving $\upmu \text{deg}$-level detection of optical rotation on the second timescale has recently been demonstrated using a frequency metrology method\cite{Bougas2022}; the chiroptical rotation is determined via the frequency splitting between the left- and right-circularly polarised cavity modes. Notably, this approach requires the laser frequency to be locked to the cavity resonance.
\begin{figure*}[htp]
    \centering
    \includegraphics[width=\textwidth]{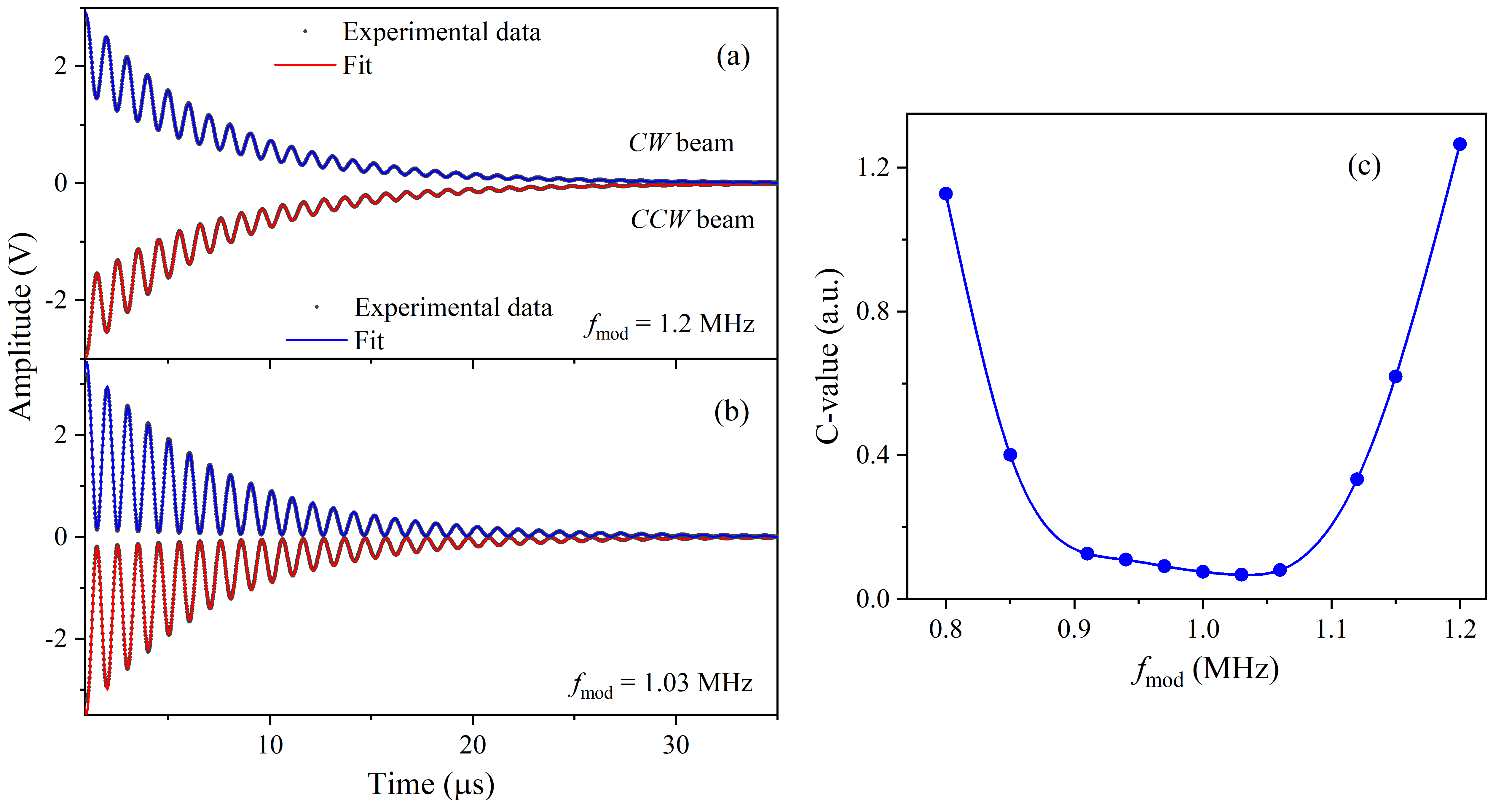}
    \caption{(a--b) Cavity ring-down signals (dark points) from the CW and CCW beams for two different modulation frequencies applied to the laser light, $f_{mod}=1.2$ and $1.03\unit{MHz}$. The red and blue solid curves are the fits of the amplitude modulation model, as given by Eq. 1, to the data. The ring-down time is $\sim6.5\unit{\upmu s}$. (c) Variation of the polarisation-modulation depth as a function of $f_{mod}$. Experimental parameters: magnetic field of $\sim0.15\unit{T}$ giving $\phi_{F}$ of $\sim3.2^{\circ}$; 2000 events averaged over 30 seconds.}
    \label{fig:modulation}
\end{figure*}
\\ \indent Most of the bow-tie cavity-based polarimeters mentioned above have employed intracavity Terbium Gallium Garnet (TGG) crystals to induce a large bias rotation or "offset" angle\cite{Sofikitis2014,Spiliotis2020,Spiliotis2020a,Bougas2015,Tran2021,Tran2021a}. However, TGG exhibits significant losses at wavelengths shorter than $600\unit{nm}$, particularly in the spectral windows below $400\unit{nm}$ (due to the $\text{Tb}^{3+}$ ${^{5}\text{D}_{3}}\leftarrow{^{7}\text{F}_{6}}$, $^{5}\text{D}_{2}$ absorption bands) and around $490\unit{nm}$ (caused by the electronic transition ${^{5}\text{D}_{4}}\leftarrow{^{7}\text{F}_{6}}$). These losses (significantly) limit the sensitivity of CRDP, and substitution for alternative magneto-optic materials is desirable. One such alternative, crystalline cerium fluoride ($\text{CeF}_{3}$) exhibits a lower absorption coefficient and a larger transparent spectral window (from $280\unit{nm}$ to $2500\unit{nm}$) than TGG\cite{Vasyliev2012,Molina2011}, making it one of the best candidates for developing highly sensitive CRDP in the UV, visible, and near-infrared regions. Indeed, recent work by Xygkis \textit{et al.}\cite{Xygkis2023} has identified $\text{CeF}_{3}$ as a Faraday-rotating crystal of high "figure of merit" for application in cavity ring-down polarimetry (on condition that the sample exhibits "ideal polishing").
\\ \indent In this paper, we first present the experimental setup for our present implementation of CRDP using a \textit{cw} laser operating at a fixed wavelength of $532\unit{nm}$ and show that external frequency modulation of the laser output at the frequency splitting between the left- and right-circularly polarised cavity modes leads to an increased detection sensitivity. We then demonstrate the high-precision measurement of the Verdet constant of crystalline $\text{CeF}_{3}$ and then present the gas-phase optical rotation measurement of enantiomers of $\upalpha$-pinene and R-($+$)-limonene. These highly precise measurements are compared with state-of-the-art computations employing the Turbomole quantum chemistry package which elucidate the optical rotatory dispersion (ORD) curves exhibited by a relevant selection of individual limonene conformers. The paper concludes with two examples of liquid-phase CRDP. Firstly, the kinetics of D-glucose mutarotation are quantified. Subsequently, the variation of the optical rotation of L-histidine with decreasing pH is presented. The latter data are interpreted with a simple thermodynamic model which allows the optical rotation of the parent zwitterion, cation and dication to be estimated, these  observations showcasing the particular utility of the apparatus for studies of aqueous solutions, and the characterisation of charge states that otherwise evade measurement.
\section{High-Precision Magnetometry with \textit{cw}-CRDP}\label{sec:Magnet}
\subsection{\textit{cw}-CRDP Apparatus}
Fig. \ref{fig:setup} shows the experimental setup for CRDP at $532\unit{nm}$. The operating principles of the cavity-enhanced continuous-wave polarimeter are explained in detail in previous publications\cite{Tran2021,Tran2021a}, and here we only highlight the necessary differences between the laser sources and the optical setups.
\begin{figure*}
    \centering
    \includegraphics[width=\textwidth]{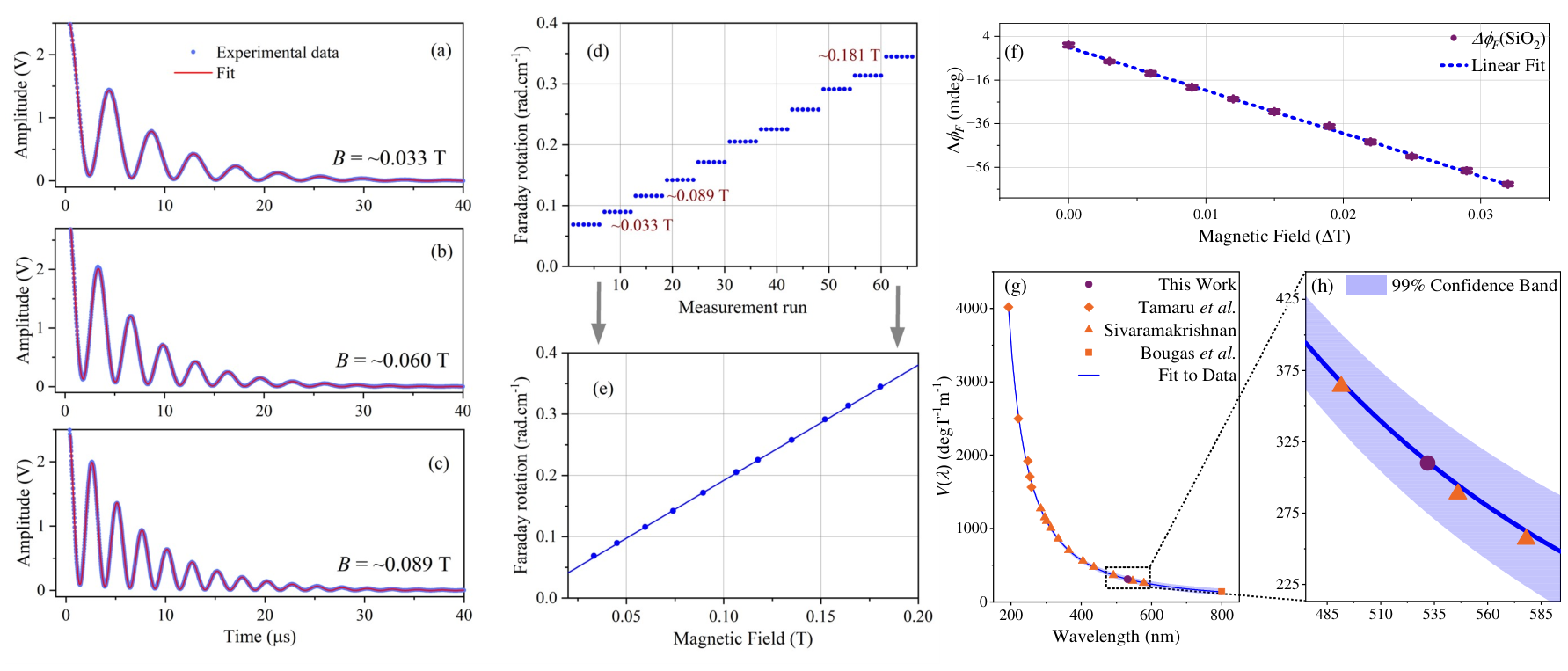}
    \caption{The Faraday effect in crystalline $\text{CeF}_{3}$ and fused silica. (a--c) Cavity ring-down signal (dark points) at magnetic fields of 0.033, 0.060, and $0.089\unit{T}$. For each recorded signal, $\sim2000$ ring-down traces are averaged in $30\unit{s}$. The red solid curves are fits of the amplitude modulation model, Eq. \ref{eq:master}, to the data. (d--e) Variation of the Faraday rotation angle as a function of magnetic field. For each $B$ field, six measurements are carried out (panel d, blue points) and the corresponding mean value is shown in panel e, (blue points). The blue line (panel e) is a linear fit to the data. (f--h) Faraday rotation measurements of a fused silica window and the corresponding Verdet dispersion calculation as compared with previous literature values from Tamaru \textit{et al.}\cite{Tamaru2021}, Sivaramakrishnan\cite{Sivaramakrishnan1956}, and Bougas \textit{et al.}\cite{Bougas2015}.}
    \label{fig:Faraday}
\end{figure*}
\\ \indent A bow-tie cavity is constructed with four high-reflectivity plano-concave mirrors (Layertec, diameter of $12.7\unit{mm}$). Each mirror has a reflectivity $\text{R}\geq99.99\%$ at $532\unit{nm}$. The total cavity length is $L=557.8\pm0.4\unit{cm}$ (width and length separations of $\sim15\unit{cm}$ and $\sim139\unit{cm}$, respectively). For scanning the cavity length, one of the cavity mirrors is mounted on a piezo-electric actuator (PZT, Piezomechanik GmBH) driven by a K-cube high-voltage piezo controller (Thorlabs, KPZ101). The polarimeter was addressed by a diode-pumped solid state (DPSS) laser at $532\unit{nm}$ (Verdi V5, Coherent) with a maximum power of $5\unit{W}$. Typically, a laser power of $\sim200\unit{mW}$ was used. The laser beam is directed to an acousto-optic modulator (AOM, AA Opto-Electronic, MT80B30-A1.5-VIS) operated at a frequency of $80\unit{MHz}$ and which has a maximum fall time of $\sim200\unit{ns}$. The AOM is driven by a radiofrequency (rf) signal generated by a voltage-controlled oscillator (VCO, AA Opto-Electronic, DRFA10Y-B-0-50.110) with a tuneable frequency range between 65 and $95\unit{MHz}$, and then power-boosted by an rf power amplifier (AA Opto-Electronic, AMPB-B-30-10.500) with a gain of $\sim34\unit{dB}$. The AOM's driving frequency can be adjusted by a voltage signal provided by a function generator (Keysight, 33509B). The VCO is also controlled by a home-made TTL trigger box which acts as the master trigger to generate cavity ring-down signals following intracavity power build-up. The resulting first-order diffraction beam from the AOM has a power of $\sim10\unit{mW}$ and is directed to the cavity. A half-waveplate ($\uplambda/2$, B-Halle) and a polarising beam-splitting cube (Thorlabs, PBS251) are used in tandem to separate the laser beam into two linearly polarised beams, p- and s-beams of clockwise (CW) and counter-clockwise (CCW) propagation, respectively. These beams are then injected into the cavity where they continue to propagate in opposing directions. Two Glan-Taylor polarisers (Thorlabs, GT10-A, extinction ratio $\geq10^{5}$) are placed at the inputs of the cavity to filter and fix the polarisation states of the incident beams. A set of lenses is used to match the Gaussian spatial profile of the CW and CCW beams with the $\text{TEM}_{00}$ mode of the cavity.
\\ \indent A \textit{c}-oriented $\text{CeF}_{3}$ crystal (E-crystal Ltd., Japan) with a diameter of $12.7\unit{mm}$ and a thickness, $l$, of $2\unit{mm}$ is inserted into one arm of the cavity to induce a large offset rotation angle, $\phi_{F}$. The crystal has been anti-reflection coated on both sides with a minimum reflectivity per surface of $\sim0.15\%$ and has a Verdet constant, $V$, of $\sim188\unit{rad\,T^{-1}\,m^{-1}}$ at $532\unit{nm}$\cite{Vasyliev2012,Molina2011,Villora2015}. This crystal is located within a magnetic field of $B<0.2\unit{T}$, provided by a pair of ring neodymium magnets (First4Magnets), and produces a large Faraday offset angle of $\phi_{F}=VBl<4^{\circ}$. Two linear polarisers (Thorlabs, LPVISC050-MP2, extinction ratio $\geq10^{5}$) are placed at the outputs of the cavity to analyse the polarisation states of the output beams. The ring-down traces are measured by applying a triangular-wave signal with a frequency of $\sim30\unit{Hz}$ to the PZT driver and scanning the cavity length. The amplitude over which the cavity length is dithered is large enough to ensure that several cavity resonances are excited by the laser. As shown in Fig. \ref{fig:setup}, the ring-down signals are detected by two photo-multiplier tubes (Hamamatsu, R928) which are driven by two high-voltage power supplies (Stanford Research, SRS PS310). The signals are then amplified by two transimpedence amplifiers (Thorlabs, TIA 60) with a bandwidth of $60\unit{MHz}$, and acquired by a 12-bit high resolution oscilloscope (Lecroy, WaveSurfer 4104HD, bandwidth of $1\unit{GHz}$, $5\unit{Gs/s}$ sampling rate) that allowed the data to be averaged (typically 2000 events over $30\unit{s}$). The detected signal is described by an exponential ring-down signal superimposed with a periodic function,
\begin{equation}\label{eq:master}
    I(t)=I_{0}\text{e}^{-t/\tau_{0}}\left[\cos^{2}\left(2\pi ft+\varphi\right)+C\right],
\end{equation}
in which $I_{0}$ is the maximum amplitude of the signal, $\tau_{0}$ is the ring-down time, $f=c\phi/2\pi L$ is the polarisation beating frequency, with $\phi$ (rad) the single-pass rotation angle, $\varphi$ the global phase offset, and $C$ accounting for any reduction in the polarisation-modulation depth. The cavity ring-down time is $\sim6.5\unit{\upmu s}$, mainly limited by the optical losses on the $\text{CeF}_{3}$ crystal resulting from both reflections and optical absorption.
\begin{figure*}
    \centering
    \includegraphics[width=\textwidth]{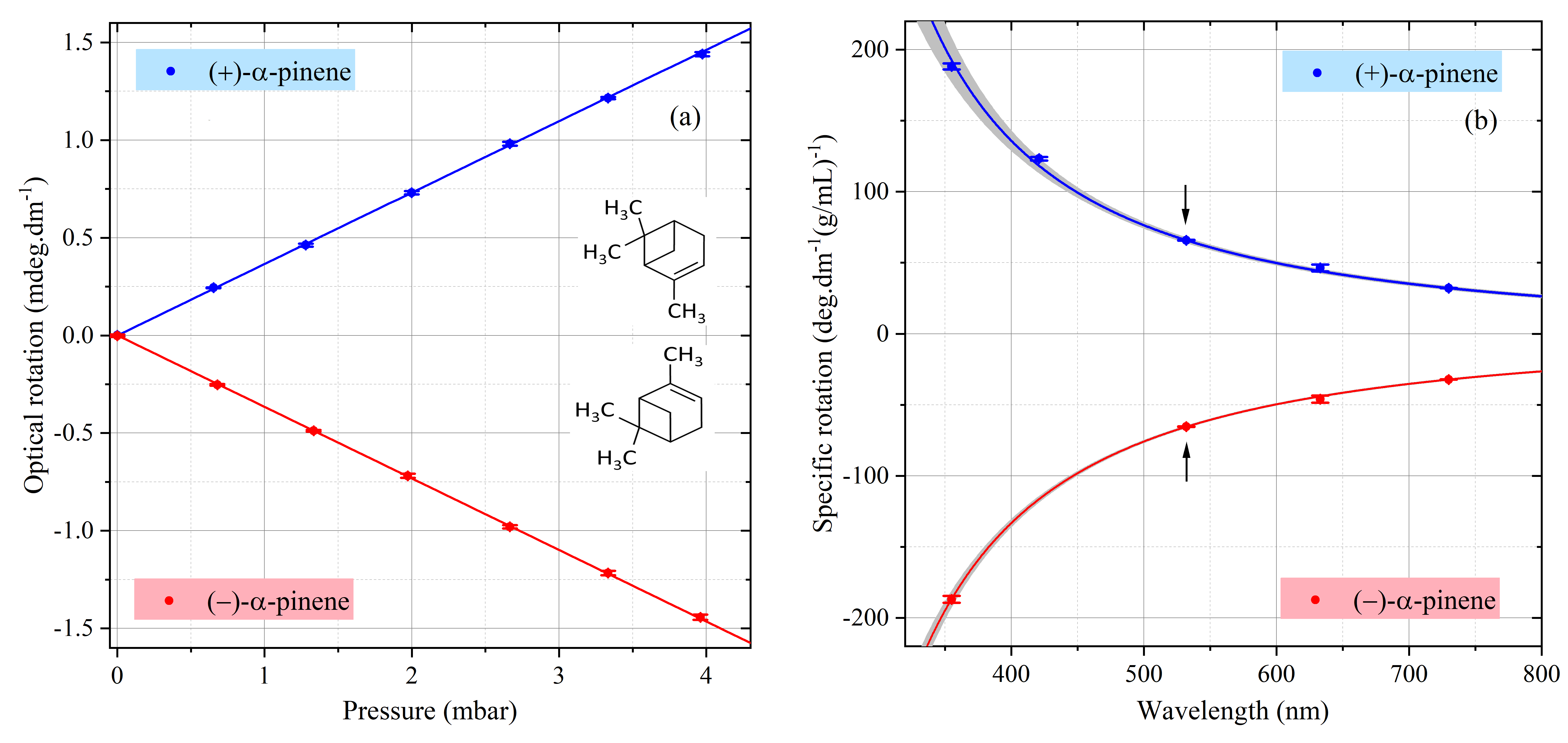}
    \caption{Variation in optical rotation of enantiomers of $\upalpha$-pinene vapour as a function of pressure. Each data point and its error bar are given by the mean and standard deviation of the 10 measurements at a given pressure, respectively. The red and blue lines are weighted linear fits to the data. Experimental parameters: $B$ field $\sim0.15\unit{T}$ ($\phi_{F}$ of $\sim3.2^{\circ}$); 2000 events averaged over $30\unit{s}$. (b) Variation of the specific rotation of ($+$)- and ($-$)-$\upalpha$-pinene as a function of wavelength (blue and red points, respectively). The black arrows indicate the optical rotation reported in this work. The grey solid curves are the fits of the wavelength dependence of the optical rotation to the data (see text for details), the grey areas representing the resulting mean prediction bands corresponding to a confidence level of 0.95.}
    \label{fig:pinene}
\end{figure*}
\subsection{Improving the Sensitivity of the Polarimeter}
Due to the Faraday effect\cite{Budker2002}, each cavity resonance is split into two circularly polarised modes, known as left- and right-resonant modes with a frequency separation of $\Delta f=c\phi_{F}/\pi L\;(\sim1\unit{MHz}$ for $\phi_{F}=3.2^{\circ}$). This separation is comparable with the bandwidth of the laser, and so the ring-down signal may be dominated by only one of these resonances, leading to a reduction in the modulation depth and thus limiting the detection sensitivity. In our recent works, we have perturbed the laser frequency by applying a Gaussian noise signal to the diode laser current, thus allowing the detection sensitivity to be improved by an order of magnitude\cite{Tran2021,Tran2021a}. For the current setup using a fixed-frequency DPSS laser, perturbation of the laser frequency can be achieved via the driving frequency of the AOM. However, the slow rise time of the AOM (\textit{ca.} $200\unit{ns}$) makes this method inefficient, and therefore we choose as an alternative, to modulate the laser frequency via the AOM at the frequency splitting between the left- and right-circularly polarised modes, $\Delta f$. This perturbs both modes equally, promoting equal, simultaneous mode excitation, and so increases the observed polarisation-modulation depth thus enhancing the CRDP detection sensitivity.
\\ \indent To study the effect of this frequency modulation of the AOM's driving frequency on the polarisation-modulation depth, we have fixed the $B$ field applied to the $\text{CeF}_{3}$ crystal at $\sim0.15\unit{T}$ (corresponding to $\Delta f\sim1\unit{MHz}$), and recorded ring-downs over a range of modulation frequencies from 0.8 to $1.3\unit{MHz}$. Fig. \ref{fig:modulation} (a--b) show exemplar ring-down signals (dark points) from the CW and CCW beams, recorded at modulation frequencies $f_{mod}=1.2$ and $1.03\unit{MHz}$, respectively. The red and blue solid curves are fits of the model given by Eq. \ref{eq:master} to the data. Fig. \ref{fig:modulation} (c) shows the variation of the polarisation-modulation depth as a function of $f_{mod}$. $C$ approaches zero when the modulating frequency approximately equals the cavity mode splitting, $\Delta f$, from 0.95 to $1.07\unit{MHz}$, allowing the detection sensitivity to be improved by around 1 order of magnitude\cite{Tran2021a}.
\subsection{Measurement of the Faraday Effect in $\text{CeF}_{3}$}
With a Verdet constant similar in magnitude to that for TGG, $\text{CeF}_{3}$ has been identified as a suitable Faraday rotator for the UV and visible regions\cite{Vasyliev2012,Molina2011,Villora2015,Xygkis2023}. In Fig. \ref{fig:Faraday} we present measurements of the Faraday effect for crystalline $\text{CeF}_{3}$ at $532\unit{nm}$. For these studies, only the CW beam is used, and its polarisation direction is rotated by a Faraday rotation angle $\phi_{F}$. The magnetic field is varied between 0.03 and $0.18\unit{T}$ by changing the distance between the two ring magnets, and is precisely measured using a factory-calibrated Hall probe magnetometer (Hirst Magnetic Instruments, GM08) with a relative accuracy of $\sim0.5\%$. For each $B$ field, the modulation frequency $f_{mod}$ has been chosen to be close to the cavity mode frequency splitting, allowing $C$ to approach zero. This was achieved via manual variation of $f_{mod}$ until maximum modulation depth had been attained. The Faraday rotation angle, $\phi_{F}=2\pi Lf/c$, with $c$ the speed of light, is determined by fitting Eq. \ref{eq:master} to the data.
\\ \indent Fig. \ref{fig:Faraday} (a--c) show sample ring-down signals (blue points) and the corresponding fits to the data (solid red curves) at fields of $B=0.033$, 0.060, and $0.089\unit{T}$. Fig. \ref{fig:Faraday} (d) shows the Faraday rotation angle as the magnetic field ranges from 0.03 to $0.18\unit{T}$. For each $B$ field, six measurements were carried out and their mean values are shown in Fig. \ref{fig:Faraday} (e) (blue points), along with a linear fit to the data. The Verdet constant of $\text{CeF}_{3}$ is determined to be $188.4\pm0.4\unit{rad\,T^{-1}\,m^{-1}}$ and is consistent with the value of $\sim188\unit{rad\,T^{-1}\,m^{-1}}$ derived from the Verdet constant dispersion model previously measured using a UV-IR spectrometer\cite{Bougas2022}, although marginally larger than the value of $171\unit{rad\,T^{-1}\,m^{-1}}$ obtained via an alternative linear, two-mirror cavity ring-down setup\cite{Xygkis2023}. 
\begin{figure*}
    \centering
    \includegraphics[width=\textwidth]{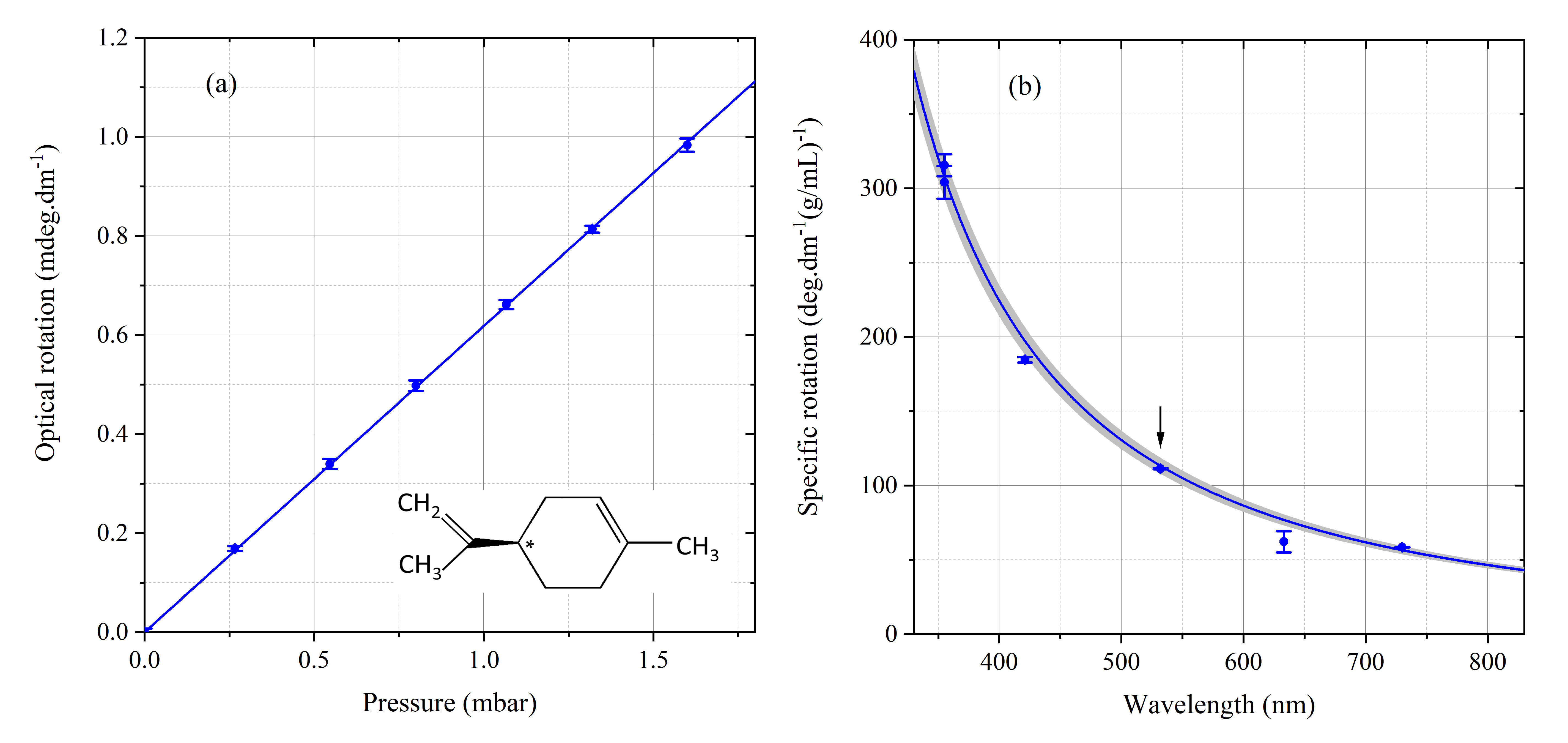}
    \caption{(a) Gas-phase optical rotation measurements of R-($+$)-limonene vapour over a range of pressures. Each data point and its error bar are given by the mean and the $2\sigma$ standard error of 10 measurements at a given pressure, respectively. The blue line is a weighted linear fit to the data. Experimental parameters: a magnetic field of $\sim0.15\unit{T}$ giving a $\phi_{F}$ of $\sim3.2^{\circ}$; 2000 events averaged over $30\unit{s}$. (b) Variation of the specific rotation of R-($+$)-limonene as a function of wavelength (blue and red points, respectively). The black arrows indicate the optical rotation of enantiomers of R-($+$)-limonene at $532\unit{nm}$ measured by our $cw$-CRDP apparatus. The grey solid curve is a fit of the wavelength dependence of the optical rotation to the data (see text for details), and the grey area is the mean prediction band resulting from the fits with a confidence level of 0.95.}
    \label{fig:limonene}
\end{figure*}
\subsection{Evaluation of the Verdet Constant of Fused Silica}
Fused silica represents another dielectric medium capable of exhibiting Faraday rotation. The material's broadband transparency and ubiquity in optical elements has motivated the characterisation of its magnetically induced circular birefringence. In a manner analogous to the aforementioned study, a cavity ring-down magnetometry analysis was conducted on a $6.35\unit{mm}$-thick fused silica window measuring 1 inch in diameter and possessing ion beam sputtered anti-reflection coatings of $0.035\%$ maximum reflectivity (Edmund Optics). In this instance, an electromagnet comprising a U-core (LEYBOLD$^{\circledR}$ 562 11), a pair of coils (LEYBOLD$^{\circledR}$ 562 13), and a brace of bored poles (LEYBOLD$^{\circledR}$ 560 31), received current over a range of 1 Ampere (R\&S$^{\circledR}$NGL202), the resulting magnetic field being estimated using a transverse Hall probe (GM08 Gaussmeter, Hirst Magnetic Instruments Ltd.). Ring-down traces were recorded at intervals of $0.1\unit{A}$---the corresponding Faraday rotation measurements are displayed in figure \ref{fig:Faraday} (f).
\\ \indent A linear fit to the Faraday rotation measurements reveals a Verdet constant of $310\pm3\unit{deg\,T^{-1}\,m^{-1}}$ at $532\unit{mn}$, in near-perfect agreement with interpolations of literature values\cite{Tamaru2021,Sivaramakrishnan1956, Bougas2015} recorded over a series of wavelengths (figures \ref{fig:Faraday} (g) and (h)). The wavelength dependence of the Verdet constant depicted in these figures was modelled using the following functional form\cite{Qiu1998, Takebe1994, Vojna2019}:
\begin{equation}\label{eq:VerdetDisp}
V(\lambda)=\frac{\alpha}{\lambda^{2}-\lambda_{0}^{2}}+\beta
\end{equation}
A fit of equation \ref{eq:VerdetDisp} to the data displayed in figure \ref{fig:Faraday} (g) returned evaluated parameter values of $(8.045\pm0.207)\times10^{7}\unit{nm^{2}\,deg\,T^{-1}\,m^{-1}}$, $8.464\pm18.020\unit{deg\,T^{-1}\,m^{-1}}$, and $130.9\pm2.0\unit{nm}$ for $\alpha$, $\beta$, and $\lambda_{0}$, respectively.
\section{Gas-Phase Measurements}
\begin{table*}
  \caption{Reported values of the specific rotation of enantiomers of $\upalpha$-pinene and R-($+$)-limonene in the gas phase. The final row presents the thermal average values at $295\unit{K}$ for R-($+$)-limonene, determined from \textit{ab-initio} calculations (see section \ref{subsec:limonene}\label{tab:pinene}).}
    \begin{center}
    \renewcommand{\arraystretch}{1.5}
        \begin{tabular}{p{0.135\textwidth}  p{0.1\textwidth}  p{0.1\textwidth}  p{0.12\textwidth}  p{0.12\textwidth}  p{0.12\textwidth}  p{0.12\textwidth}  p{0.12\textwidth} }
            \toprule
            \toprule
            \multirow{2}{0.14\textwidth}{Optically active sample}&\multirow{2}{*}{Assay purity}&\multirow{2}{0,1\textwidth}{\centering Optical purity ($ee$)}&\multicolumn{5}{c}{Optical rotation ($\unit{deg\,dm^{-1}\,(g/ml)^{-1}}$)}\\
             \cmidrule{4-8}
            &&& \centering $355\unit{nm}$\cite{Muller2000}& 
 \centering $421\unit{nm}$\cite{Bougas2022}& \centering  $532\unit{nm}$& \centering $633\unit{nm}$\cite{Wilson2005} & \hspace{0.4cm} $730\unit{nm}$\cite{Tran2021,Tran2021a} \\
             \hline
             ($+$)-$\upalpha$-pinene& \centering $99\%$ &\centering $97\%$ &$\phantom{-}188.2\pm2.2$&$\phantom{-}123.2\pm1.3$&$\phantom{-1}65.81\pm0.30$& $\phantom{-}46.3\pm2.5$&$\phantom{-}32.10\pm0.13$\\
             ($-$)-$\upalpha$-pinene&\centering $99\%$&\centering $97\%$&$-187.0\pm2.4$&\;&$\phantom{,}-65.90\pm0.20$&$-46.0\pm2.5$&$-32.21\pm0.11$\\
             \multirow{2}{*}{R-($+$)-limonene}&\multirow{2}{*}{\hspace{0.55cm} $97\%$}&\multirow{2}{*}{\hspace{0.6cm}$98\%$}&$\phantom{-}315.5\pm7.4$&\multirow{2}{*}{$\phantom{-}184.6\pm1.8$} & \multirow{2}{*}{$\phantom{-}110.49\pm0.47$}&\multirow{2}{*}{$\phantom{-}62.1\pm7.1$}&\multirow{2}{*}{$\phantom{-}59.83\pm0.25$} \\
            &&&$\phantom{-}304.2\pm11\phantom{.}$&&&&\\
             \hline
             R-($+$)-limonene$^{\dagger}$ &\centering $-$&\centering $-$&\hspace{0.25cm}370.8&$\phantom{-}$239.6$^{\ddagger}$&$\phantom{-}$139.3&$\phantom{-}$94.9&$\phantom{-}$69.8\\
             \bottomrule
             \bottomrule
        \end{tabular}
      	\begin{flushleft}
		\footnotesize{
			$^{\dagger}$Calculation at $295\unit{K}$\\
            $^{\ddagger}$Interpolated value at $421\unit{nm}$ from data in Figure \ref{fig:Newman}
		}	
	\end{flushleft}
    \end{center}
\end{table*}
The experimental apparatus described in section \ref{sec:Magnet} has been used to perform measurements of gas-phase optical activity. If a chiral sample is introduced into one arm of the cavity, producing a chiroptical rotation, $\phi_{C}$, then after each cavity roundtrip the polarisation states of the CW and CCW beams are rotated by an amount $\phi_{CW}=\phi_{F}+\phi_{C}$ and $\phi_{CCW}=\phi_{F}-\phi_{C}$, respectively. By fitting the ring-down traces to Eq. \ref{eq:master}, the oscillating frequencies, $f_{CW}=c\left(\phi_{F}+\phi_{C}\right)/2\pi L$ and $f_{CCW}=c\left(\phi_{F}-\phi_{C}\right)/2\pi L$ can be determined, and thus the chiroptical rotation is found to be
\begin{equation}\label{eq:chiropt}
    \phi_{C}=\frac{\pi L}{c}\left(f_{CW}-f_{CCW}\right).
\end{equation}
\indent To measure the optical rotation of gas phase chiral molecules, a $1.1\unit{m}$-long gas cell was inserted into one arm of the cavity. The cell was enclosed by two $1\unit{inch}$ diameter, $6.35\unit{mm}$ thick fused silica windows (Edmund Optics, ion beam sputtered AR coated with $\text{R}<0.035\%$). Each window was placed in a home-made mount and rested on two O-rings to minimise forces acting on the window, thus reducing birefringence. The gas pressure in the cell was measured by a capacitance pressure gauge (Leybold, CTR100, pressure range of 100 Torr). The cell was evacuated below $10^{-6}\unit{mbar}$ using a compact turbo-molecular pump station (Pfeiffer, HiCube 300). Chiral samples were contained in a reservoir and their vapour injected into the gas cell via a dosing valve. A $B$ field of $\sim0.15\unit{T}$ was applied to the $\text{CeF}_{3}$ crystal, corresponding to a Faraday rotation angle of $\phi_{F}\sim3.2^{\circ}$. The empty cavity ring-down times for both beams are $\sim4.7\unit{\upmu s}$, corresponding to $\sim255$ round trips and an effective path length through the gas cell of $\sim280\unit{m}$.
\subsection{Gas-Phase Optical Rotation of ($+$)- and ($-$)-$\upalpha$-Pinene}
Fig. \ref{fig:pinene} (a) shows the gas-phase optical rotation per unit length produced by the enantiomers of $\upalpha$-pinene (Sigma-Aldrich, enantiomeric excess $ee=97\%$) in the pressure range 0--$4\unit{mbar}$. Due to a combination of scattering and absorption of the intracavity light by $\upalpha$-pinene, the ring-down time decreases from $4.7\unit{\upmu s}$ to $4.2\unit{\upmu s}$ as the pressure increases from 0 to $4\unit{mbar}$. Ten measurements were carried out at each pressure with $f_{CW}$ and $f_{CCW}$ determined by fitting data to Eq. \ref{eq:master}, and $\phi_{C}$ then determined using Eq. \ref{eq:chiropt}. The uncertainty in $\phi_{C}$ for each measurement is calculated from the fitting errors of $f_{CW}$ and $f_{CCW}$. There is a slight difference between $f_{CW}$ and $f_{CCW}$ even without the present of a chiral sample. This effect is due to the small inhomogeneities in the magnetic field and the slightly different propagation of the CW and CCW beams through the $\text{CeF}_{3}$ crystal, and potentially in addition to the presence of non-planarity-induced optical rotation (NPI-OR), a phenomenon described and quantified in work by Bougas and co-workers\cite{Bougas2015}. We therefore introduce a frequency offset $\delta=f_{CW}-f_{CCW}$ to Eq. \ref{eq:chiropt}; $\delta$ is $\sim1\unit{kHz}$ and does not change during a set of measurements. Values for $\phi_{C}$ and its errors are the weighted mean and $2\sigma$ weighted standard error of the ten measurements at each pressure respectively. The optical rotation precision is $\sim30\unit{\upmu deg}$ per cavity pass over a $30\unit{s}$ detection period, as given by the error bar for $\phi_{C}$. This corresponds to a time-averaged sensitivity of $\sim164\unit{\upmu deg\,Hz^{-1/2}}$, and is $\sim3$ times better than compared to previous pulsed CRDP investigations at $532\unit{nm}$\cite{Spiliotis2020} but less than that of the $cw$-CRDP approach reported at $730\unit{nm}$\cite{Tran2021a} ($\sim10\unit{\upmu deg}$ per cavity pass). This is due to the larger optical losses incurred by use of an intra-cavity $\text{CeF}_{3}$ crystal in the present setup. The red and blue solid lines are linear weighted fits to the data and possess gradients $\text{d}\phi_{C}/\text{d}p=0.3655\pm0.0015\unit{mdeg\,dm^{-1}\,mbar^{-1}}$ for ($+$)-$\upalpha$-pinene, and $-0.3660\pm0.0009\unit{mdeg\,dm^{-1}\,mbar^{-1}}$ for ($-$)-$\upalpha$-pinene. The gas-phase specific optical rotation, $\left[\alpha\right]^{295\unit{K}}_{532\unit{nm}}$, is thus determined to be $65.81\pm0.30\unit{deg\,dm^{-1}\,(g/ml)^{-1}}$ for ($+$)-$\upalpha$-pinene and $-65.45\pm0.20\unit{deg\,dm^{-1}\,(g/ml)^{-1}}$ for ($-$)-$\upalpha$-pinene at a wavelength of $532\unit{nm}$ and a temperature of $22^{\circ}\text{C}$, where the quoted uncertainty includes a contribution related to the lengths of the intracavity gas cell and the cavity itself, the magnitude of which corresponds to $\sim0.11\unit{deg\,dm^{-1}\,(g/ml)^{-1}}$.
\\ \indent To our knowledge, these are the first measurements of the gas-phase optical rotations of enantiomers of $\upalpha$-pinene at $532\unit{nm}$. Previously, similar gas-phase optical rotation measurements on $\upalpha$-pinene ($ee\sim97\%$) enantiomers have been demonstrated at $355\unit{nm}$ and $633\unit{nm}$ by Vaccaro \textit{et al.} at $25^{\circ}\text{C}$\cite{Muller2000,Wilson2005}, and at $730\unit{nm}$ by our group at $22^{\circ}\text{C}$\cite{Tran2021,Tran2021a}. The rotation of nominally enantiomerically pure ($+$)-$\upalpha$-pinene at $421\unit{nm}$ and $21^{\circ}\text{C}$\cite{Bougas2022} has also recently been reported. These measurements are summarised in Table \ref{tab:pinene}. It is noted that the gas-phase optical rotation of ($+$)- and ($-$)-$\upalpha$-pinene at $800\unit{nm}$ has also been reported by Sofikitis \textit{et al.}, with reported rotations of $17.57\pm0.57$ and $-18.04\pm0.98\unit{deg\,dm^{-1}\,(g/ml)^{-1}}$, respectively. The enantiomeric excess of the samples has not, however, been provided\cite{Sofikitis2014}.
\\ \indent Fig. \ref{fig:pinene} (b) shows the measured rotatory powers of the two $\upalpha$-pinene enantiomers between 355 and $730\unit{nm}$. Assuming that the optical rotation is dominated by the intense $\pi\rightarrow\pi^{*}$ transition at around $200\unit{nm}$, the ORD curve can be modelled as
\begin{equation}\label{eq:ORD}
    \left[\alpha\right]^{T}_{\lambda}=\frac{A_{T}}{\left(\lambda^{2}-\lambda_{0}^{2}\right)}
\end{equation}
where $A_{T}$ is the temperature-dependent amplitude and $\lambda_{0}$ is the resonant wavelength of the electronic transition that produces the chiroptical rotation\cite{Vojna2019}. The blue and red solid curves are weighted fits of this simple model to the data. Fig. \ref{fig:pinene} (b) also shows the mean prediction band resulting from the fits with a confidence level of 0.95 (grey areas). The resulting closest UV transition and the optical rotation amplitude are found to be $\lambda_{0}=205\pm2\unit{nm}$ ($210\pm2\unit{nm}$ when including the reported measurement at $421\unit{nm}$) and $A_{T}=\left(1.575\pm0.006\right)\times10^{7}$ ($\left(1.571\pm0.019\right)\times10^{7}$ including measurement at $421\unit{nm}$) $\unit{deg\,dm^{-1}\,(g/mL)^{-1}\,nm^{2}}$ for ($+$)-$\upalpha$-pinene, and $\lambda_{0}=203\pm2\unit{nm}$ and $A_{T}=\left(1.583\pm0.005\right)\times10^{7}\unit{deg\,dm^{-1}\,(g/mL)^{-1}\,nm^{2}}$ for ($-$)-$\upalpha$-pinene.
\subsection{Gas-Phase Optical Rotation of R-($+$)-Limonene}\label{subsec:limonene}
\begin{figure}
    \centering
    \includegraphics[width=0.45\textwidth]{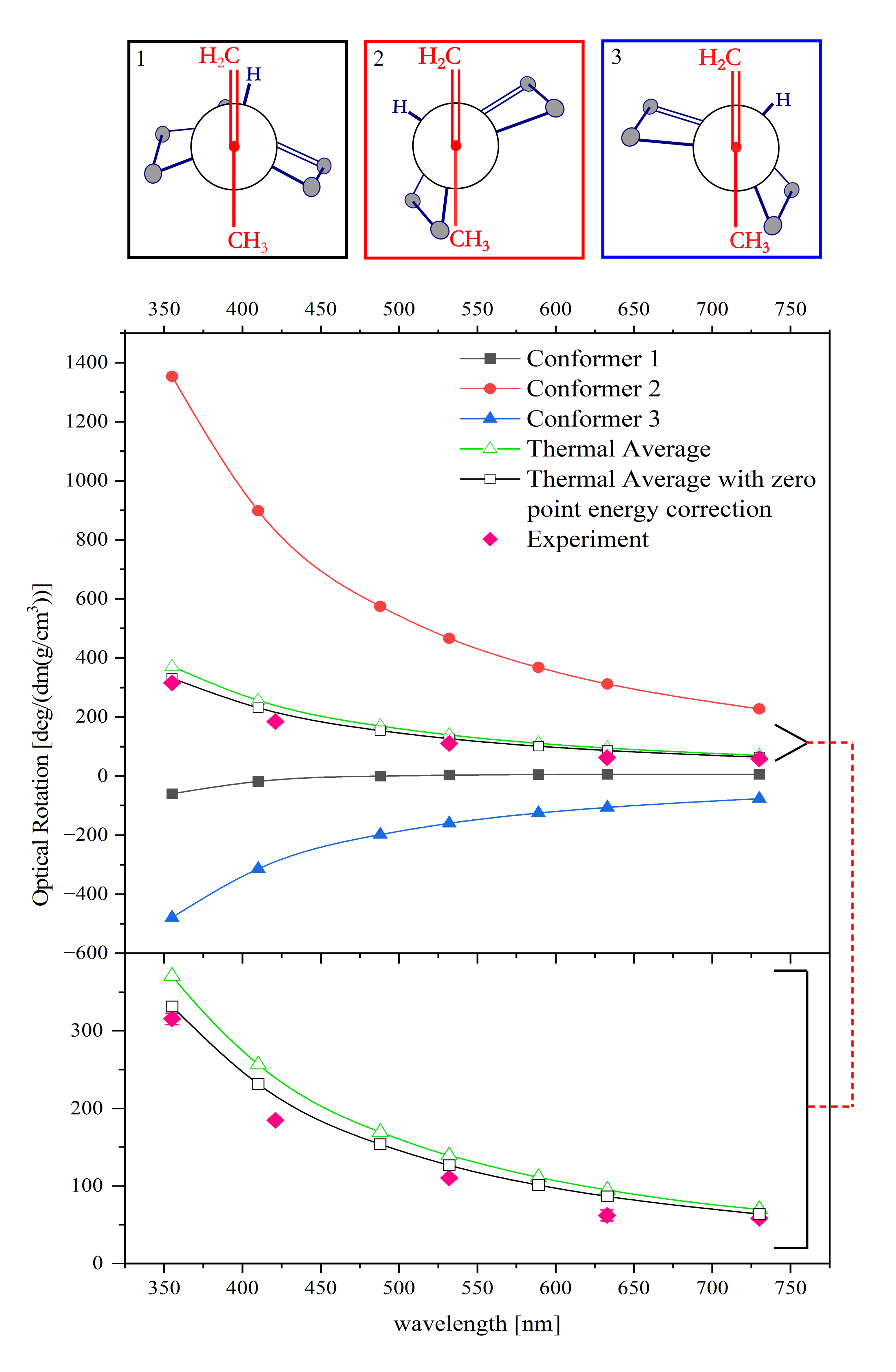}
    \caption{Three distinct conformers of limonene depicted as Newman projections (upper panes) and associated \textit{ab-initio} ORD curves (lower graph). The $295\unit{K}$ thermal averages (with and without zero-point energy corrections) are shown as well as the experimental results collated in Table \ref{tab:pinene} (pink diamonds). The lower pane is a zoom of the thermal averages and experimental results}
    \label{fig:Newman}
\end{figure}

The gas-phase chiroptical power of R-($+$)-limonene (Sigma-Aldrich, $ee=99\%$) has also been quantified and Fig. \ref{fig:limonene} (a) shows its optical rotation per unit length as a function of pressure between 0 and $1.6\unit{mbar}$. Due to a combination of optical absorption and scattering, $\tau$ is observed to decrease from $4.7\unit{\upmu s}$ for zero sample pressure to $\sim2.5\unit{\upmu s}$ at $1.6\unit{mbar}$; the error bar for each data point correspondingly increasing with pressure. The linear weighted fit to the data has gradient $\text{d}\phi_{C}/\text{d}p=\left(0.6147\pm0.0024\right)\unit{mdeg\,dm^{-1}\,mbar^{-1}}$ from which the optical rotation is determined to be $\left(110.49\pm0.47\right)\unit{deg\,dm^{-1}\,(g/mL)^{-1}}$ at $532\unit{nm}$ and $22^{\circ}\text{C}$. The quoted uncertainty of the specific rotation includes a systematic uncertainty relating to the intracavity gas-cell and cavity lengths. There is, to our knowledge, only one other gas-phase optical rotation measurement of R-($+$)-limonene (assay$\geq95\%$, unknown enantiomeric excess) at $532\unit{nm}$\cite{Spiliotis2020} with a reported specific rotation of $\sim70\unit{deg\,dm^{-1}\,(g/mL)^{-1}}$. Measurements of the optical rotation of R-($+$)-limonene vapour have been performed at 355 and $633\unit{nm}$ at $25^{\circ}\text{C}$\cite{Muller2000,Wilson2005}, at $421\unit{nm}$ at $21^{\circ}\text{C}$\cite{Bougas2022}, and $730\unit{nm}$ at $22^{\circ}\text{C}$\cite{Tran2021a}. The findings of these studies are summarised in Table \ref{tab:pinene} and displayed in Fig. \ref{fig:limonene} (b). Noting that there are also weak transitions at 174.5, 210, 218 and \textit{ca.} $400\unit{nm}$ in limonene\cite{Brint1984,Fujiki2013}, if we assume that the optical rotation is dominated by the intense $\pi\rightarrow\pi^{*}$ transition of the cyclohexane moiety at $186\unit{nm}$, then the simple model given by Eq. \ref{eq:ORD} can be used to describe the ORD curve of R-($+$)-limonene. The blue solid line in Fig. \ref{fig:limonene} (b) is then the weighted fit to the data for which the electronic transition is fixed at $\lambda_{0}=186\unit{nm}$. Within this crude estimation, the resulting optical rotation amplitude is found to be $A_{T}\sim\left(2.81\pm0.05\right)\times10^{7}\unit{deg\,dm^{-1}\,(g/mL)^{-1}\,nm^{2}}$. Once again, the grey area encompassing the blue line represents a $95\%$ confidence level.
\\ \indent Limonene is a cyclic monoterpene and at room temperature can exist in three distinct conformers (rotamers) defined by a relative moiety rotation about a single bond such that eclipsing interactions between the ring and isopropenyl moiety are minimised. The different conformers (denoted 1 to 3, with increasing enthalpy) are depicted as Newman projections in Fig. \ref{fig:Newman} (upper panes) and have relative energies 0, 0.44, and $1.91\unit{kJ\,mol^{-1}}$, respectively. These energies were obtained from CCSD(T)(F12*)/aug-cc-pVQZ single point energy calculations\cite{Raghavachari:CPL157-479,Hattig:JCP132-231102,Franzke:JCTC,Dunning:JCP90-1007} at B3LYP/def2-TZVPP optimised geometries, accounting for vibrational zero-point energy at the B3LYP/def2-TZVPP level of theory.\cite{Becke:JCP98-1372,Stephens:JPC98-11623,Weigend:PCCP7-3297}
{ Including a correction for the zero-point energy yields relative energies of 0, 0.89 and 2.28$\unit{kJ\,mol^{-1}}$ for the three conformers, which shifts the thermal average slightly, as shown in Fig. \ref{fig:Newman}.}
ORD values for each conformer were computed using linear response\cite{Furche:TCC-2005} with the CAM-B3LYP functional\cite{Yanai:CPL393-51} in the length representation with the aug-cc-pVQZ basis set.\cite{Dunning:JCP90-1007}
The calculations reveal that each of the conformers exhibit a different optical rotatory power, the interplay of which dictates the calculated (and experimentally observed) ORD curves shown in Fig. \ref{fig:Newman}. Notably, two of the three conformers (1 and 3) display optical rotations that are smaller in magnitude and of the opposite sense to the other (2); at room temperature ($295\unit{K}$) the population-weighted rotation for conformer 2 dominates. The thermal average rotation is shown by the green unfilled triangles in Fig. \ref{fig:Newman} and is in reasonable agreement with experiment.
\\ 
\section{Liquid-Phase Measurements}\label{sec:liquids}
\begin{figure*}
    \centering
    \includegraphics[width=\textwidth]{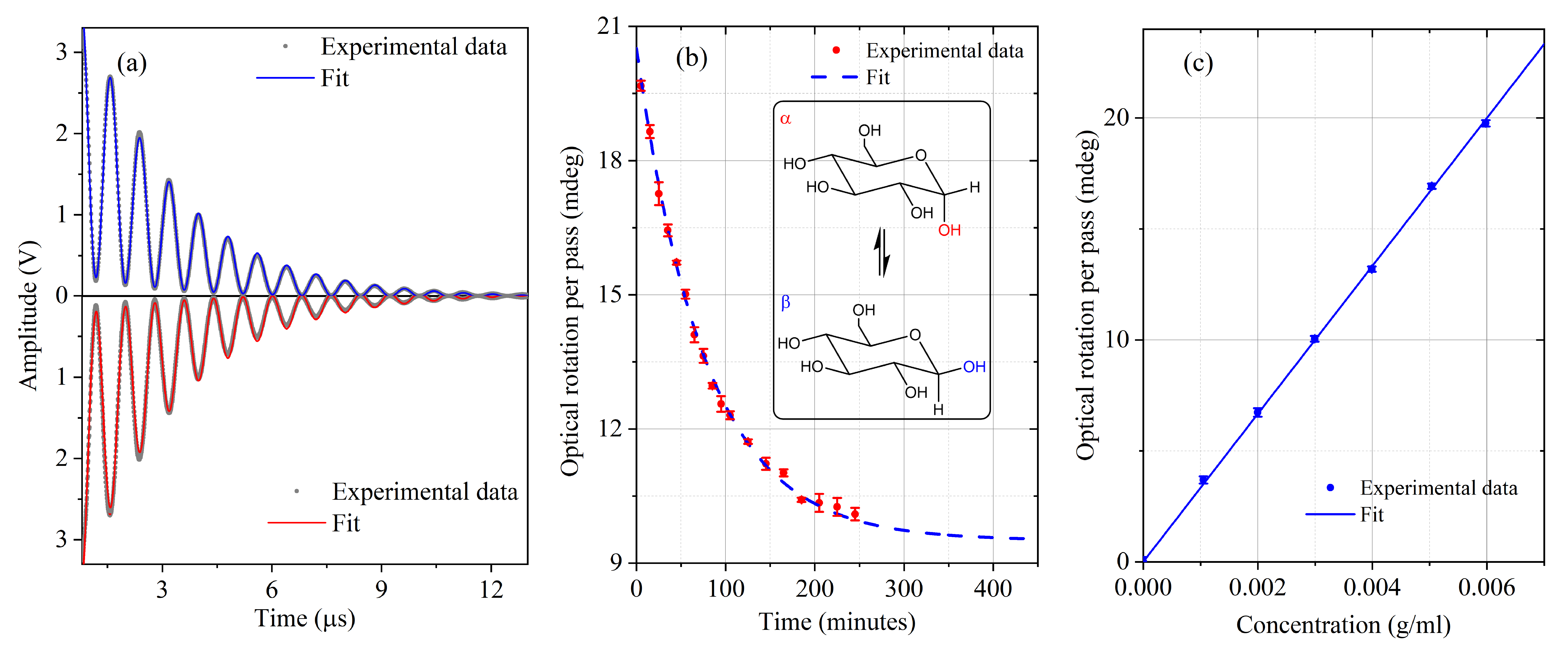}
    \caption{Liquid-phase optical rotation measurements of chiral aqueous samples. (a) Cavity ring-down signals (dark points) from the CW and CCW beams probing liquid water. The red and blue solid curves are fits of Eq. \ref{eq:master} to the data. $\tau$ is $\sim2.5\unit{\upmu s}$. Experimental parameters: $B$ field $\sim0.15\unit{T}$ ($\phi_{F}\sim3.2^{\circ}$); 2000 events averaged over $30\unit{s}$. (b) Variation of rotation angle per round-trip pass in the cavity for $0.003\unit{g/mL}$ D-glucose as a function of time (red points). Each data point and its error are the average and standard error of 5 data sets, respectively. The dashed blue curve is a weighted exponential fit to the data. (c) Variation of optical rotation angle per round-trip pass in the cavity as a function of D-glucose concentration (after 5 hours waiting time). Each data point and its error are the average and standard error of 10 data sets at the given concentration, respectively. The blue line is a weighted linear fit to the data.}
    \label{fig:glucose}
\end{figure*}
For measuring the chiroptical properties of aqueous samples, a home-made $5.3\unit{mm}$-long flow cell was inserted into one arm of the cavity (see Fig. \ref{fig:setup}). The cell consists of two fused silica windows (Layertec GmBH) which are AR coated for the air/fused-silica interface ($\text{R}<0.1\%$) and non-coated for the water/fused-silica interface. These windows are separated by a FFKM O-ring (Polymax, thickness of $5.30\pm0.03\unit{mm}$) and mounted on two kinematic cage-compatible mounts (Thorlabs, KC1/M). Chiral aqueous solutions were injected into the flow cell using a syringe pump (Harvard Apparatus, 11) with a flow rate of $\sim0.2\unit{mL\,min^{-1}}$. Once again, a $B$ field of $\sim0.15\unit{T}$ was applied to the $\text{CeF}_{3}$ crystal ($\phi_{F}\sim3.2^{\circ}$). Fig. \ref{fig:glucose} (a) shows typical ring-down signals observed with the flow cell filled with water. Each signal is an average of 2000 events over 30 seconds. $\tau$ is $\sim3\unit{\upmu s}$ and corresponds to $\sim135$ round trips in the cavity and an effective path length through the flow cell of $\sim0.72\unit{m}$.
\subsection{Mutarotation of D-Glucose}
\begin{figure*}
    \centering
    \includegraphics[width=\textwidth]{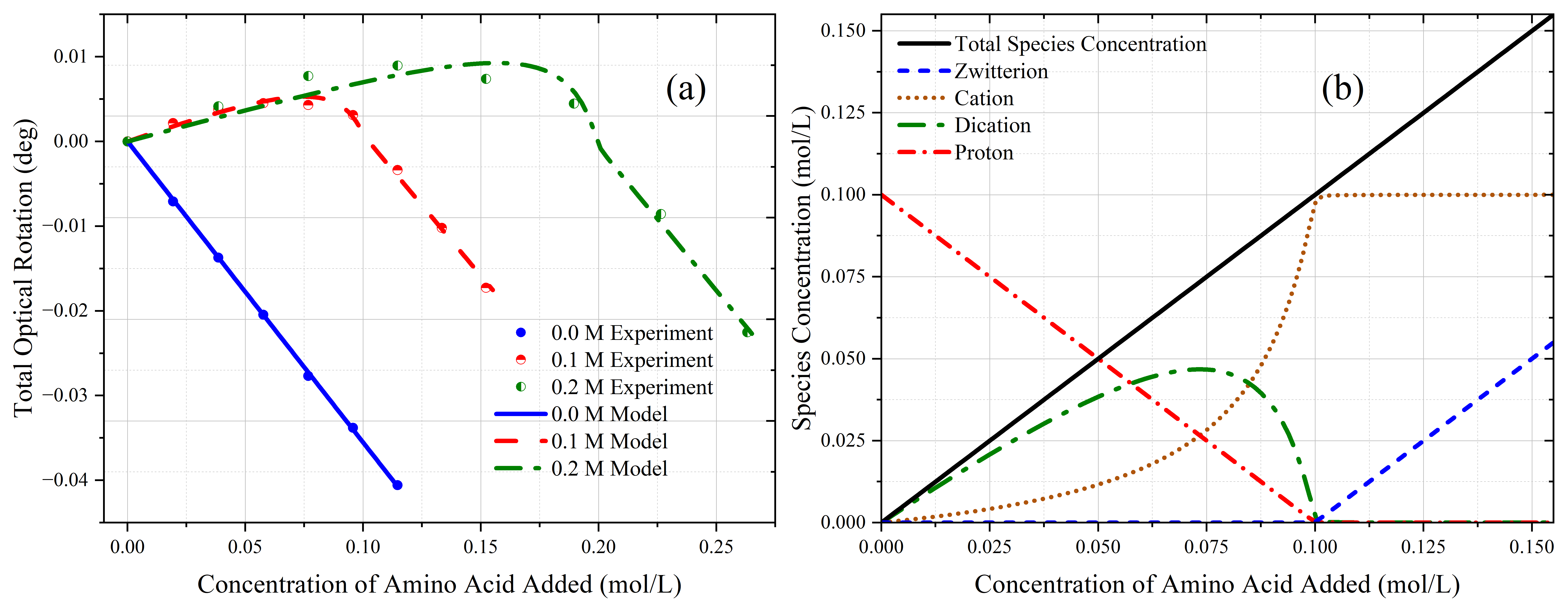}
    \caption{pH-dependent optical rotation of L-histidine. (a) Experimental findings for optical rotation of L-histidine under three acidity regimes compared with modeled values (see text for details). (b) Representative calculated concentration profile for L-histidine charge states in solution, alongside the predicted "free proton" concentration, in the presence of $0.1\unit{M}$ HCl$_\text{(aq.)}$.}
    \label{fig:histidine}
\end{figure*}
We use this apparatus to probe the mutarotation of D-glucose via measurement of its specific rotation---where the dissolved chiral molecules eventually equilibrate between different anomers. Mutarotation manifests as a time dependent change in the specific rotation (as the two anomers in this case have different specific rotations), and Fig. \ref{fig:glucose} (b) shows the time evolution of the optical rotation angle for $3\times10^{-3}\unit{g/mL}$ D-glucose in water over a period of 250 minutes. For each time point, 5 measurements were carried out over a total observation duration of $75\unit{s}$. Each data point and its error bar in Fig. \ref{fig:glucose} (b) are given by the weighted mean and the standard error of the five measured rotation angles, respectively. The reaction kinetics are well described by a simple, first-order exponential model $\phi_{C}(t)=\phi_{C,\;t\rightarrow\infty}+A\text{e}^{-kt}$ (dashed curve). Fitting this to the data reveals a rate constant for mutarotation of 
$k=1.298\pm0.047\times10^{-2}\unit{min^{-1}}$ at $18^{\circ}\text{C}$, in good agreement with recent broadband Mueller ellipsometry measurements conducted by Vala \textit{et al.}\cite{Vala2023}.
\\ \indent Fig. \ref{fig:glucose} (c) shows the specific rotation of D-glucose in water as the mutarotation approaches completion ($\sim5$ hours observation period). Each data point and its error are the average and the standard error of 10 data sets at a given concentration, respectively. The optical rotation of D-glucose is determined to be $62.96\pm0.38\unit{deg\,dm^{-1}\,(g/mL)^{-1}}$ at $532\unit{nm}$, where the quoted uncertainty of the specific rotation includes a systematic uncertainty relating to the intracavity gas cell and cavity lengths.
\subsection{pH-Dependent Optical Rotation of L-Histidine}
The apparatus has also been used to investigate the optical activity of chiral amino acids and the variation of the same as a function of the solution environment. As an example, we present data describing how the optical rotation of L-histidine varies as a function of both its own concentration as well as the solution acidity. The results are discussed in light of the well documented Clough-Lutz-Jirgensons (CLJ) rule\cite{Greenstein1961}. Finally, a thermodynamic analysis is conducted in order to ascertain the concentration profiles of the various ionic species in solution, from which the optical rotations of the latter can be approximated. The results of this calculation thus provide a quantitative interpretation of the observed CLJ behaviour.
\\ \indent The total optical rotation of L-histidine was measured as a function of amino acid concentration under three environments: 0.0, 0.1, and $0.2\unit{M}$ $\text{HCl}_{\text{(aq.)}}$. The results are shown in Fig. \ref{fig:histidine} (a). The CLJ rule, namely that the acidification of an aqueous solution of an L-amino acid causes a positive increase in its optical rotation\cite{Greenstein1961}, is clearly observed and is consistent with previous experimental findings\cite{Hayashi1966}. In order to interpret the data presented in Fig. \ref{fig:histidine} (a), concentration profiles of the chiral species in solution, each of which contributes to the observed rotation, were calculated. This was achieved by solving the below system of equations over the amino acid concentration range ($\left[\text{AA}_{\text{tot}}\right]$) for the three environments studied, in which $\left[\text{H}^{+}\right]$ denotes the concentration of "free protons", i.e., hydroxonium ions, in solution. It must be noted that this represents only a preliminary investigation, and neglects electrostatic contributions and previously-documented phenomena such as specific anion effects\cite{Rossi2007}.
\begin{gather}
    \left[\text{AAH}^{+}\right]K_{a_{1}}-\left[\text{AA}^{\pm}\right]\left[\text{H}^{+}\right]=0 \label{eq:catKa}\\
    \left[\text{AAH}^{2+}_{2}\right]K_{a_{2}}-\left[\text{AAH}^{+}\right]\left[\text{H}^{+}\right]=0 \label{eq:dicatKa}\\
    \left[\text{AAH}^{+}\right]+\left[\text{AAH}^{2+}_{2}\right]+\left[\text{H}^{+}\right]-\left[\text{HCl}\right]=0 \label{eq:neut}\\
    \left[\text{AA}^{\pm}\right]+\left[\text{AAH}^{+}\right]+\left[\text{AAH}^{2+}_{2}\right]-\left[\text{AA}_{\text{tot}}\right]=0 \label{eq:mass}
\end{gather}
\\ \indent Under the experimental conditions used, it can be assumed that only the zwitterion, cation and dication of L-histidine could be present in significant quantities in solution. In the above equalities, the concentrations of these species are represented as $\left[\text{AA}^{\pm}\right]$, $\left[\text{AAH}^{+}\right]$ and $\left[\text{AAH}^{2+}_{2}\right]$, respectively. Of these equations, \ref{eq:catKa} and \ref{eq:dicatKa} represent the acid dissociation equilibria of the cation and dication, respectively, where $\text{p}K_{a_{1}}$ (corresponding to the imidazole side chain of the histidine molecule) was taken to be 6.00 while $\text{p}K_{a_{2}}$ (representing the carboxylic acid moiety) was set equal to 1.82\cite{Carey2013}. Eq. \ref{eq:neut} accounts for electroneutrality in the bulk solution, with \ref{eq:mass} describing the system's mass balance. The results of one such set of concentration profile evaluations are depicted in Fig. \ref{fig:histidine} (b).
\\ \indent With the concentration profiles in hand, a sequential least squares programming (SLSQP) method, fitting to the experimental findings, was employed to determine the optimised values for the individual optical rotations. This was conducted for all three environments simultaneously under the assumption that the optical rotation of an individual species does not change dramatically between each condition set. The results of these calculations are displayed in Table \ref{tab:histidine}. The total optical rotation is a concentration-weighted sum of the contributions from each species in solution, and so the optimised individual rotations can be used to infer the predicted total observed rotation as a function of amino acid concentration at each acidity regime. This analysis was conducted and the results are depicted in Fig. \ref{fig:histidine} (a).
\begin{table}
    \begin{center}
    \renewcommand{\arraystretch}{1.5}
        \begin{tabular}{l r r r}
            \toprule
            \toprule
            Optically active\;\;\;\;\;&\multicolumn{3}{c}{Optical Rotation ($\text{deg}\,\text{dm}^{2}/\text{mol}$)} \\
            \cmidrule{2-4}
            sample&This work&\;\;\;\;\;\;\;\;\;Theory$^{\text{a,b}}$&\;\;\;Experimental\cite{Greenstein1961} \\
            \midrule
            L-His$^{\pm}$&$-6.693$&$-6.24,-4.57$&$-5.98$ \\
            L-His$^{+}$&$0.095$&-,-&- \\
            L-His$^{2+}$&$1.533$&$10.54,8.83$&$1.83$ \\
             \bottomrule
             \bottomrule
        \end{tabular}
        \caption{Predicted optical rotation values for L-histidine species in solution. $^{\text{a.b}}$Results of time-dependent density functional theory calculations using B3LYP (a) and BHLYP (b) functionals\cite{Kundrat2008}.\label{tab:histidine}}
    \end{center}
\end{table}
\\ \indent It is noteworthy that this simple, first-principles methodology shows quite good agreement with both theoretical\cite{Kundrat2008} and experimental\cite{Greenstein1961} values, some of which are also displayed in Table \ref{tab:histidine}. For example, at $532\unit{nm}$ we predict an $8.23\unit{deg\,dm^{2}\,mol^{-1}}$ rotation increase from zwitterion to dication, agreeing well with Greenstein's experimental finding of $7.81\unit{deg\,dm^{2}\,mol^{-1}}$ at $589.3\unit{nm}$. Notably, this analysis provides access to approximate optical rotations of intermediate charge states, in this case the monocation, otherwise difficult to extract directly from experimental observations.
\section{Conclusions}
We have developed a new variant of continuous-wave CRD polarimetry using a fixed-wavelength laser source at $532\unit{nm}$. The sensitivity of this apparatus is optimised via frequency modulation of the laser output to regularly and simultaneously excite both the non-degenerate left- and right-circularly polarised cavity modes. The CRDP technique has been demonstrated by evaluating the Verdet constants of $\text{CeF}_{3}$ and fused silica, as well as the gas-phase optical rotation of ($+$)- and ($-$)-$\upalpha$-pinene, and (R)-($+$)-limonene. The limit of precision for the gas-phase optical rotation measurements is $\sim30\unit{\upmu deg}$ per cavity pass, limited by the optical losses on the $\text{CeF}_{3}$ crystal, and the uncertainty on the specific optical rotation is better than  $0.3\unit{deg\,dm^{-1}\,(g/mL)^{-1}}$. The measurements of limonene are interpreted via state-of-the-art computational evaluations of the optical rotations of the three thermally accessible conformers of this species in the gas phase. Their relative energies were determined to span a range of approximately $2\unit{kJ\,mol^{-1}}$. In addition, the CRDP technique has been extended to measurements in the liquid phase with a limit of detection of $\sim120\unit{\upmu deg}$ per cavity pass, corresponding to a time-averaged sensitivity of $\sim657\unit{\upmu deg\,Hz^{-1/2}}$. Specifically, the mutarotation of D-glucose has been tracked over 4 hours and a rate constant for mutarotation of $k=1.298\pm0.047\times10^{-2}\unit{min^{-1}}$ at a temperature of $18^{\circ}\text{C}$ was determined. Finally, the optical rotation of L-histidine has been determined as a function of the total amino acid concentration, and the acidity of the solution. These measurements allowed estimation of the optical rotatory powers of the aqueous L-histidine zwitterion, cation and dication, thus presenting a promising means of evaluating the chiroptical properties of otherwise unobservable charge states in solution. The results of these final investigations also provide accessible, empirical quantification of the time-tested Clough-Lutz-Jirgensons rule for amino acids.

%
%

%

\begin{acknowledgments}
This work was funded by the European Commission Horizon 2020, ULTRACHIRAL Project (grant no. FETOPEN, ID no. 737071).
\end{acknowledgments}
\section*{Data Availability}
The data that support the findings of this study are available from the corresponding author upon reasonable request.
\section*{References}
\bibliography{UltraChiral532nm}
\end{document}